\documentclass[12pt]{article}
\usepackage{graphicx}

\newcommand\pubnumber{SNSN-xxx-xx} 
\newcommand\pubdate{\today}

\def\lbabar{\mbox{{\large\sl B}\hspace{-0.4em} {\normalsize\sl A}\hspace{-0.03em}{\large\sl B}\hspace{-0.4em} {\normalsize\sl A\hspace{-0.02em}R}}}
\usepackage{relsize}
\def\babar{\mbox{\slshape B\kern-0.1em{\smaller A}\kern-0.1em
    B\kern-0.1em{\smaller A\kern-0.2em R}}}

\def\napoli{Department of Physics\\
Universit\'e de Montr\'eal, Montr\'eal, CANADA}
\def\support{\footnote{On behalf of the \babar\ Collaboration}}

\def\Title#1{\begin{center} {\Large #1 } \end{center}}
\def\Author#1{\begin{center}{ \sc #1} \end{center}}
\def\Address#1{\begin{center}{ \it #1} \end{center}}

\newcommand\pubblock{\rightline{\begin{tabular}{l} \pubnumber\\
         \pubdate  \end{tabular}}}
\newenvironment{Abstract}{\begin{quotation}  }{\end{quotation}}
\newenvironment{Presented}{\begin{quotation} \begin{center} 
             PRESENTED AT\end{center}\bigskip 
      \begin{center}\begin{large}}{\end{large}\end{center} \end{quotation}}

\def\beq{\begin{equation}}
\def\eeq#1{\label{#1}\end{equation}}
\def\eeqn{\end{equation}}

\def\beqa{\begin{eqnarray}}
\def\eeqa#1{\label{#1}\end{eqnarray}}
\def\eeqan{\end{eqnarray}}

\let\bar=\overbar

\def\Dslash{\not{\hbox{\kern-4pt $D$}}}
\def\dslash{\not{\hbox{\kern-2pt $\del$}}}

\def\BR{\mbox{\rm BR}}

\def\msb{{\bar{\ssstyle M \kern -1pt S}}}

\newcommand{\DBR}{\mbox {$\Delta \ensuremath{\cal B}$}}
\newcommand{\gevsq}{\ensuremath{\mathrm{\,Ge\kern -0.1em V^2\!}}}
\newcommand{\gev}{\ensuremath{\mathrm{\,Ge\kern -0.1em V\!}}}
\def\BR {\ensuremath{\cal B}}
\def\ulnu{$b \rightarrow u \ell \nu$}
\def\pilnu{$B^{0} \rightarrow \pi^{-} \ell^{+} \nu$}

\def\etaplnu{$B^{+} \rightarrow \eta^{\prime} \ell^{+} \nu$}

\def\bfpilnu{${\BR}(B^{0} \rightarrow \pi^{-} \ell^{+} \nu)$}
\def\bfetalnu{${\BR}(B^{+} \rightarrow \eta \ell^{+} \nu)$}
\def\bfetaplnu{${\BR}(B^{+} \rightarrow \eta^{\prime} \ell^{+} \nu)$}
\def\bfrholnu{${\BR}(B^{0} \rightarrow \rho^{-} \ell^{+} \nu)$}
\def\bfpilnuqq  {$\DBR(q^2)$}
\def\mes        {\mbox{$m_{\rm ES}$}}
\def\DeltaE     {\mbox{$\Delta E$}}

\def\qq{$q^2$}
\def\BFval{$\left(1.42 \pm 0.05_{stat} \pm 0.07_{syst} \right) \times 10^{-4}$}
\def\BFetaval{$\left(0.36 \pm 0.05_{stat} \pm 0.04_{syst} \right) \times 10^{-4}$}
\def\BFetapval{$\left(0.24 \pm 0.08_{stat} \pm 0.03_{syst} \right) \times 10^{-4}$}

\def\VubFpzVal{$\left(8.6 \pm 0.3_{stat} \pm 0.3_{syst} \right) \times {10^{-4}}$}
\def\BFvalslac{$\left(1.41 \pm 0.05_{stat} \pm 0.07_{syst} \right) \times 10^{-4}$}
\def\BFrhoval{$\left(1.75 \pm 0.15_{stat} \pm 0.27_{syst} \right) \times 10^{-4}$}
\def\VubFpzValslac{$\left(10.8 \pm 0.6 \right) \times {10^{-4}}$}

\begin{document}
\begin{titlepage}
\pubblock

\vfill
\Title{Studies of exclusive charmless semileptonique $B$ decays\\
and extraction of $|V_{ub}|$ at \lbabar}
\vfill
\Author{ Martin Simard\support}
\Address{\napoli}
\vfill
\begin{Abstract}
We report on recent measurements of the branching fractions (BFs) for the decay channels
$B^0\rightarrow\pi^-\ell^+\nu$, $B^+\rightarrow\pi^0\ell^+\nu$, 
$B^+\rightarrow\eta\ell^+\nu$, $B^+\rightarrow\eta^{\prime}\ell^+\nu$,
$B^0 \rightarrow\rho^-\ell^+\nu$ and $B^+ \rightarrow\rho^0\ell^+\nu$.
We obtain very precise values of the total branching fractions for these decays,
as well as partial branching fractions as a function of $q^2$ for the decay channels
$B^0 \rightarrow\pi^-\ell^+\nu$, $B^+\rightarrow\eta\ell^+\nu$ 
and $B^0 \rightarrow \rho^-\ell^+\nu$.
In particular, we use the partial branching fractions of the $B^0 \rightarrow\pi^-\ell^+\nu$
decay channel and form-factor calculations to extract several values of $|V_{ub}|$.
When we compared these values of $|V_{ub}|$ to the one measured in inclusive semileptonic $B$ decay,
we find that two of them are consistent, within large theoretical uncertainties.
\end{Abstract}
\vfill
\begin{Presented}
Proceedings of CKM2010,
the 6th International Workshop on the CKM Unitarity Triangle\\
University of Warwick, UK\\
6-10 September 2010
\end{Presented}
\vfill
\end{titlepage}
\def\thefootnote{\fnsymbol{footnote}}
\setcounter{footnote}{0}

\section{Introduction}
  
Semileptonic decays are best to measure $|V_{ub}|$~\cite{CKM} because they are much easier to understand theoretically than hadronic 
decays and they are far more abundant than leptonic decays.
In particular, the decay rate of $B\rightarrow\pi\ell\nu$ is proportional to $|V_{ub}f_+(q^2)|^2$,
where $f_+(q^2)$ is the theoretical form-factor calculation as a function of $q^2$,
the momentum transferred squared.
A precise value of $|V_{ub}|$ from the exclusive $B\rightarrow\pi\ell\nu$ decay
can be performed to test the QCD calculations and to constrain the description of
weak interactions and CP violation in the Standard Model.

We present two recent analyses in \babar\ where $|V_{ub}|$ is obtained in the study 
of the exclusive $B\rightarrow\pi\ell\nu$ decay. 
In the $\pi-\eta$ analysis~\cite{Simard}, we study three decay modes: 
$B^0\rightarrow\pi^-\ell^+\nu$, $B^+\rightarrow\eta\ell^+\nu$ and 
$B^+\rightarrow\eta^{\prime}\ell^+\nu$. 
In the $\pi-\rho$ analysis~\cite{Jochen}, we study four decay modes: 
$B^0\rightarrow\pi^-\ell^+\nu$, $B^+\rightarrow\pi^0\ell^+\nu$, 
$B^0\rightarrow\rho^-\ell^+\nu$ and $B^+ \rightarrow\rho^0\ell^+\nu$.
These decays involve a $b \rightarrow u$ transition via the coupling of a $W$ gauge boson. 
The value of $|V_{ub}|$ can be extracted from the partial branching fraction, 
$\DBR(q^2)$, measured as a function of $q^2$:
\begin{equation}
\label{eqDz}
|V_{ub}| = \sqrt{\frac{\DBR(q^2)}{\tau_{B^0}\Delta\zeta(q^2)}},~~~
\Delta\zeta(q^2) =  \frac{G_F^2}{24\pi^{3}} \int_{q^2_{min}}^{q^2_{max}} 
|\vec{p}_{\pi}|^3|f_{+}(q^2)|^2 dq^2,
\end{equation} 
where $\tau_{B^0} = 1.525 \pm 0.009$ ps~\cite{PDG10} is the $B^0$ lifetime. 
As can be seen in Eq. \ref{eqDz}, $\Delta\zeta(q^2)$ depends on the $f_{+}(q^2)^2$ 
form factor that is provided by QCD calculations from light cone sum rules at low
$q^2$ values and lattice QCD at high $q^2$ values. 



\section{Experimental method}

\begin{table}
\begin{center}
\begin{tabular}{lp{0.1cm}cp{0.1cm}cp{0.1cm}c}
\hline\hline
Analysis            & &$\pi-\eta$ & &$\pi-\rho$       \\ \hline
Luminosity on $\Upsilon(4S)$ peak & & 422.6 $fb^{-1}$ & & 349.0 $fb^{-1}$  \\
Number of $B\bar{B}$ pair events& & 464 millions & & 377 millions     \\   
$q^2$ evaluation    & & $(P_B-P_{meson})^2$ & & $(P_{\ell}+P_{\nu})^2$ \\ 
Cut strategy        & & cuts, $q^2$ dependent& &NN, $q^2$ dependent   \\
Cut selection       & & loose $\nu$ cuts  & &tighter $\nu$ cuts       \\
Signal efficiency   & & 8\% to 15\% & & 6\% to 7\%                    \\
Background/signal   & &  11.5       & & 6.3                           \\
$B^0 \rightarrow \pi^-\ell^+\nu$ yield & & $11778\pm435$ & & $10604\pm376$ \\
Number of $q^2$ bins in $\pi$ mode & &  12 & & 6                      \\
Systematic uncertainties & & full gaussian & & $\pm 1 \sigma$         \\
\hline\hline
\end{tabular}
\caption[]{\label{compare} Comparison of various characteristics for the two
recent analyses in \babar.} 
\end{center}
\end{table}

  The main differences between the $\pi-\eta$ and the $\pi-\rho$ analyses 
are summarized in Table~\ref{compare}. 
Both analyses use an untagged technique. This means that only one of the
$B$ mesons of the $B\bar{B}$ pair 
is reconstructed. 
Values of $q^2$ are determined using different methods which do 
lead to some variations in their values.
The two-dimensional distribution of true versus reconstructed values of $q^2$ yields a detector 
response matrix which is used to unfold the measured $q^2$ distribution onto the
true $q^2$ one. 


Backgrounds can be broadly grouped into three main categories: decays 
arising from \ulnu\ transitions, decays in other 
$B\bar{B}$ events and decays in continuum events. 
Given the sufficient number of events in the $\pi-\eta$ analysis for the $\pi^-\ell^+\nu$ 
decay mode, the data samples can be subdivided in $12$ bins of $q^2$ for the 
signal.
For the $\pi-\rho$ analysis and the other decay modes, a smaller number of events leads
us to restrict the number of bins used in the fit.
We use the \DeltaE-\mes\ histograms, obtained from the Monte Carlo (MC) simulation,
as two-dimensional probability density functions (PDFs),
to extract the yields of the signal and backgrounds as a function of $q^2$ in our fit to the data. 

In each analysis, the systematic uncertainties are estimated from the variations
of the resulting partial BF values when
the data are re-analyzed with different simulation parameters.
In the $\pi-\eta$ analysis, for each parameter, we produce
new PDFs by varying 
randomly the parameter value over a complete gaussian 
distribution whose standard deviation is given by the parameter uncertainty.
One hundred such variations are done for each parameter. 
The systematic uncertainty of a parameter is given by its
RMS value of the resulting partial BF distribution from these one hundred variations. 
In the $\pi-\rho$ analysis, the systematic uncertainties 
are evaluated by $\pm 1 \sigma$ variation for each parameter.


\section{Results}



\begin{figure}
 \begin{minipage}[t]{.48\linewidth}
\centering\includegraphics[width=\linewidth, height=7cm]{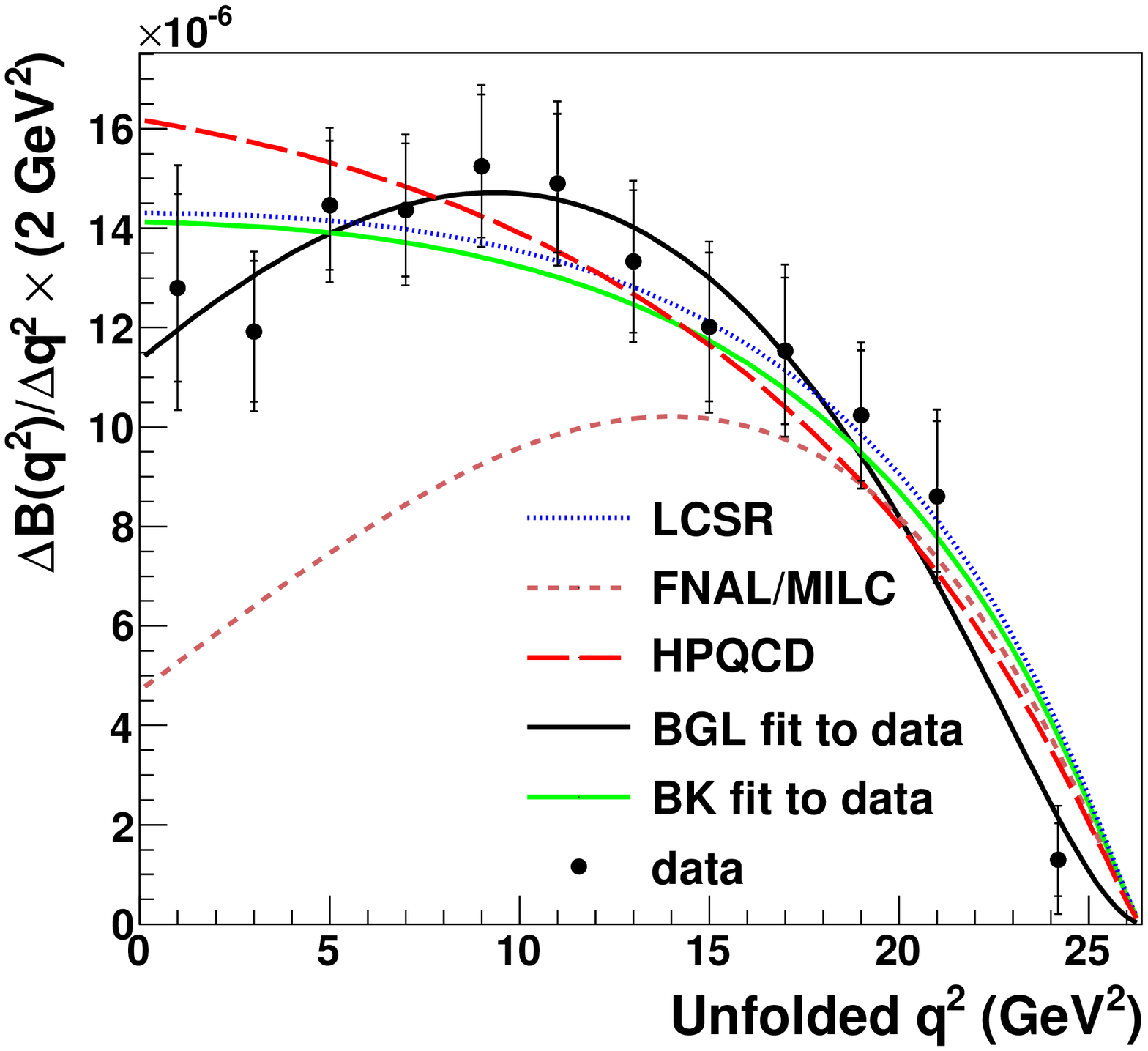}
\caption[]{\label{figFplus} Partial \bfpilnuqq\ spectrum in 12 bins of \qq\ for 
\pilnu\ decays in the $\pi-\eta$ analysis. 
The solid green and black curves show the result of the fit to the data of the 
BK~\cite{BK} and BGL~\cite{BGL} parametrizations, respectively. The data are also
compared to unquenched LQCD calculations (HPQCD~\cite{HPQCD06}, FNAL~\cite{FNAL})
and an LCSR calculation~\cite{LCSR}.}
 \end{minipage} \hfill
 \begin{minipage}[t]{.48\linewidth}
\centering\includegraphics[width=\linewidth, height=7cm]{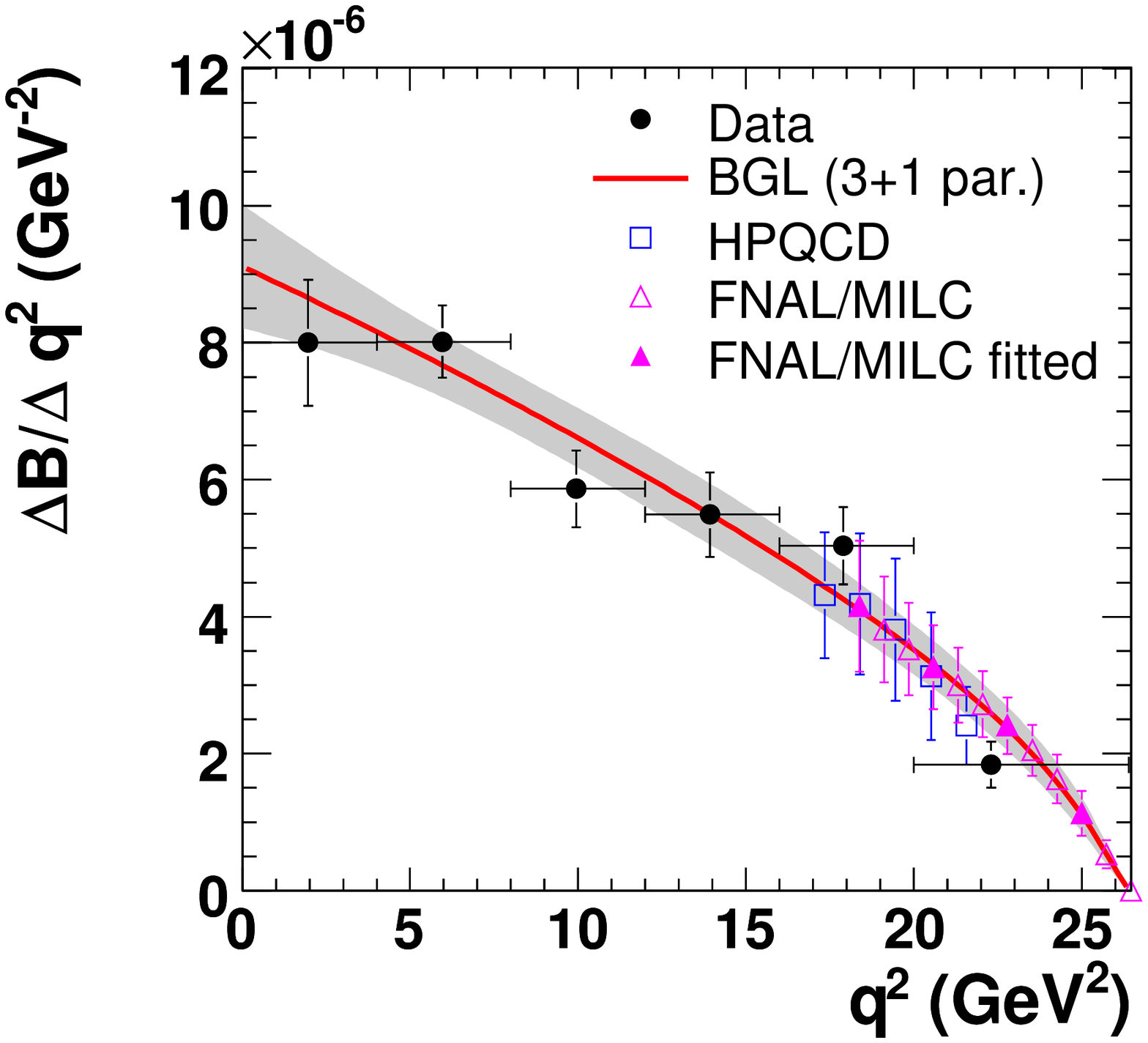}
\caption[]{\label{slacfigFplus} Partial $B\rightarrow\pi\ell\nu$ spectrum in 6 bins of \qq\ 
for $B\rightarrow\pi\ell\nu$ decays in the $\pi-\rho$ analysis. The red curve represents the 
simultaneous fit to the data points and four theoretical points produced with the
FNAL LQCD calculation (magenta, closed triangles).} 
 \end{minipage}
\end{figure}

  The experimental \bfpilnuqq\ distributions for \pilnu\ decays are displayed in 
Fig.~\ref{figFplus} in the $\pi-\eta$ analysis, together with two 
parametrizations and three QCD calculations, and in Fig~\ref{slacfigFplus} in 
the $\pi-\rho$ analysis, combining the charged and neutral pion channels assuming isospin symmetry. 
From the BGL expansion extrapolated to \qq\ $= 0$, 
we obtain the value of $|V_{ub}f_{+}(0)|$ in both analysis as shown in Table~\ref{vubtable}.
These values
differs from both analysis since the experimental  
distributions do look different. However, the individual values of the partial 
branching fractions are indeed consistent with each other for the two analyses.
The comparison between theory and experiment in their \qq\ ranges of validity 
shows that all three QCD calculations are compatible with the data as we can see in Table~\ref{vubtable}.
\newpage

  Since, in the $\pi-\rho$ analysis, the number of data points is limited to two 
above $q^2$ = 16 \gevsq, it was deemed desirable to undertake a simultaneous fit of theoretical and experimental 
points to extract a value of $|V_{ub}|$. 
We obtain the total BFs \bfpilnu\ = \BFval, \bfetalnu\ = \BFetaval\ and 
\bfetaplnu\ = \BFetapval\ in the $\pi-\eta$ analysis, and \bfpilnu\ = \BFvalslac\
and \bfrholnu\ = \BFrhoval\ in the $\pi-\rho$ analysis. Values of $|V_{ub}|$ 
obtained in our two analyses are given in Table~\ref{vubtable}. They range from 
$(3.0 - 3.8)\times 10^{-3}$. 


\begin{table*}

\begin{center}
\begin{tabular}{ccc}
\hline\hline
  Analysis       & $\pi-\eta$ &$\pi-\rho$  \\ \hline
HPQCD~\cite{HPQCD06} ($q^2 > 16$ \gevsq) & 
$3.24\pm 0.13\pm 0.16{}^{+0.57}_{-0.37}$ & $3.21 \pm 0.17{}^{+0.55}_{-0.36}$ \\
FNAL~\cite{FNAL} ($q^2 > 16$ \gevsq) & $3.14\pm 0.12\pm 0.16{}^{+0.35}_{-0.29}$ &
$2.95 \pm 0.31$ \\
LCSR~\cite{LCSR} ($q^2 < 12$ \gevsq) & $3.70\pm 0.07\pm 0.09{}^{+0.54}_{-0.39}$ &
$3.78\pm 0.13{}^{+0.55}_{-0.40}$\\
$|V_{ub}f_{+}(0)|$ & \VubFpzVal\ & \VubFpzValslac\ \\
\hline\hline
\end{tabular}
\end{center}
\caption[]{\label{vubtable} Values of $|V_{ub}|\times10^{-3}$ derived from the 
form-factor calculations for the \pilnu\ decays. 
}
\end{table*}

\section{Summary}

 \indent
  It is estimated that there is less than 20\% overlap in the selected event 
samples between the two analyses for the \pilnu\ decay 
channel. It is thus very satisfying to note that there is excellent agreement 
between the results of the two analyses. The values of the total BFs obtained in 
our work are the most precise total BFs to date. Our value of the total BF for 
\etaplnu, with a significance of $3.0\sigma$, is an order of magnitude smaller 
than the CLEO result~\cite{CLEOpilnu2}. 
The three values of $|V_{ub}|$ are all acceptable according to the data. Two of 
them~\cite{HPQCD06,LCSR} are consistent, within large theoretical uncertainties, 
with the value measured in inclusive semileptonic $B$ decays: $|V_{ub}| = (4.27 
\pm 0.38) \times {10^{-3}}$~\cite{PDG10}.


\end{document}